\documentclass[fleqn,11pt]{article}

\usepackage{amsfonts,amsmath,amssymb}
\usepackage{latexsym,color,cite}

\usepackage[utf8]{inputenc}
\usepackage{graphicx} 

\def\onecol{\onecolumn \mathindent 2em}

\def\noi{\noindent}

\newcommand{\Title}[1]{\noi {{\Large\bf #1}}\\[1ex]}

\def\Aunames#1{\noi{\bf #1}}
\def\au#1{${}^{#1}$}
\def\Addresses#1{\medskip\noi \protect
	\begin{description}\itemsep -3pt {\it #1} \end{description}}
\def\adr#1#2{\item[${}^{#1}$]{\it #2}}

\newcommand{\Abstract}[1]{\vskip 2mm \begin{center}
        \parbox{16.4cm}{\small\noi #1} \end{center}\medskip}

\def\email#1#2{\footnotetext[#1]{e-mail: #2}\addtocounter{footnote}{1}}

\def\nq{\hspace*{-1em}}
\def\nqq{\hspace*{-2em}}

\def\qq{\qquad}

\def\lal{&&\nq{}}
\def\eq{Eq.\,}
\def\eqs{Eqs.\,}
\def\beq{\begin{equation}}
\def\eeq{\end{equation}}
\def\bear{\begin{eqnarray}}
\def\bearr{\begin{eqnarray} \lal}
\def\ear{\end{eqnarray}}
\def\earn{\nonumber \end{eqnarray}}

\def\nnn{\nonumber\\ \lal }
\def\nnnv{\nonumber\\[5pt] \lal }

\def\yyy{\\[5pt] \lal }

\def\e{{\,\rm e}}
\def\D{\partial}

\def\diag{\mathop{\rm diag}\nolimits}

\def\const{{\rm const}}
\def\Half{\dfrac 12}

\def\then{\ \Rightarrow\ }

\def\eqn#1{\eq\eqref{#1}}
\def\rf{\eqref}
\def\mn{_{\mu\nu}}
\def\MN{^{\mu\nu}}
\def\mN{_\mu^\nu}

\def\M{{\mathbb M}}
\def\R{{\mathbb R}}

\def\cF{{\mathcal F}}
\def\cK{{\mathcal K}}
\def\cL{{\mathcal L}}
\def\cO{{\mathcal O}}

\def\sph{spherically symmetric}
\def\ssph{static, spherically symmetric}
\def\bh{black hole}
\def\bhs{black holes}
\def\wh{wormhole}
\def\whs{wormholes}

\def\emag{electromagnetic}
\def\Scw{Schwarz\-schild}
\def\RN{Reiss\-ner-Nord\-str\"om}
\def\Areg{A_{\rm reg}}
\def\rreg{r_{\rm reg}}
\begin{document}
\onecol

\Title{A regular center instead of a black bounce}

\Aunames{S. V. Bolokhov,\au{a;1} K. A. Bronnikov,\au{a,b,c;2} and M. V. Skvortsova\au{a;3}}

\Addresses{\small
\adr a {Institute of Gravitation and Cosmology, Peoples' Friendship University of Russia (RUDN University),\\ 
		ul. Miklukho-Maklaya 6, Moscow 117198, Russia}
\adr b {Center for Gravitation and Fundamental Metrology, VNIIMS, 
		Ozyornaya ul. 46, Moscow 119361, Russia}
\adr c {National Research Nuclear University ``MEPhI'', 
		Kashirskoe sh. 31, Moscow 115409, Russia}
        }
        
\Abstract
  {The widely discussed ``black-bounce'' mechanism of removing a singularity at $r=0$ in a spherically 
  symmetric space-time, proposed by Simpson and Visser, consists in removing the point $r=0$ and
  its close neighborhood, resulting in emergence of a regular minimum of the spherical radius 
  that can be a wormhole throat or a regular bounce. Instead, it has been recently proposed to make 
  $r=0$ a regular center by properly modifying the metric, still preserving its form in regions far from 
  $r=0$. Different algorithms of such modifications have been formulated for a few classes of singularities. 
  The previous paper considered space-times whose Ricci tensor satisfies the condition $R^t_t =R^r_r$,
  and regular modifications were obtained for the Schwarzschild, \RN\ metics, and two examples of solutions 
  with magnetic fields obeying nonlinear electrodynamics (NED). The present paper considers regular
  modifications of more general space-times, and as examples, modifications with a regular center have 
  been obtained for the Fisher (also known as JNW) solution with a naked singularity and a family of 
  dilatonic black holes. Possible field sources of the new regular metrics are considered in the framework 
  of general rlativity (GR), using the fact that any static, spherically symmetric metric with a combined 
  source involving NED and a scalar field with some self-interaction potential. This scalar field is, in general, 
  not required to be of phantom nature (unlike the sources for black bounces), but in the examples 
  discussed here, the possible scalar sources are phantom in a close neighborhood of $r=0$ and are
  canonical outside it.
   }

\email 1 {boloh@rambler.ru}
\email 2 {kb20@yandex.ru} 
\email 3 {milenas577@mail.ru}

\section{Introduction}

  The classical theories of gravity, beginning with general relativity (GR), in many of their most important 
  solutions predict the emergence of space-time singularities that designate natural applicability limits of these 
  theories. Since one can hardly believe in the reality of such formal inferences as infinite densities or curvatures,
  there have been many attempts to avoid such singularities thus making the theories universally applicable,
  still remaining in the framework of classical rather than quantum notions of space-time. Among them 
  one can recall both low-energy or low-curvature limits of various models of quantum gravity 
  \cite{QG1, QG2, QG3, kelly20, achour20, kunst20, ash20a, bambi13, gingrich24, koshelev24, bueno24}
  and application of geometries more general than Riemannian, including torsion, nonmetricity etc.
  or using gravitational theories more general than GR still using Riemannian geometry. 
   
  On the same trend, significant popularity was gained by simple phenomenological models of 
  space-time that modify and regularize the Riemannian metric behavior in very strong field regions but 
  almost preserve the geometry in the regions of weak or moderate curvature. This approach was probably 
  used for the first time by J. Bardeen in 1968 \cite{bardeen68}, who proposed to consider, instead of the 
  \Scw\ line element, the regular \bh\ metric 
\beq            \label{bardeen}
			ds^2 = \bigg(1 - \frac{2Mr^2}{(r^2 + a^2)^{3/2}}\bigg)dt^2 
			- \bigg(1 - \frac{2Mr^2}{(r^2 + a^2)^{3/2}}\bigg)^{-1} dr^2 	- r^2 d\Omega^2,
	\qq		d\Omega^2 = d\theta^2 + \sin^2\theta d\varphi^2.
\eeq  
  This metric is defined in the range $r \geq 0$. At small values of the parameter $a$ relative 
  to the \Scw\  mass $M$, which looks most natural ($a < a_{\rm crit}\approx 0.77 M$), it describes a \bh\ 
  with two horizons, at $a = a_{\rm crit}$ there is a single extremal horizon, and at $a > a_{\rm crit}$
  horizons are absent, and we observe a soliton-like, or star-like geometry, and with any values of $a$
  there is a regular center at the origin $r=0$. Similar properties are observed in Hayward's metric \cite{hayward}
  that differs from Bardeen's by the replacement $(r^2 + a^2)^{3/2} \mapsto r^3 + 2 M \ell^2$, $\ell = \const$.  
  
  A great number of regular field models have been considered since then, even if we restrict our attention to
  \ssph\ configurations.(see, e.g., \cite{reg1, reg2, reg3} for recent reviews).
  Still, if our interest is not in regular geometries obtained directly but in regular models
  constructed, like \rf{bardeen}, from singular ones by introducing a certain regularization parameter, we must 
  above all pay attention to the metric proposed by Simpson and Visser (SV) \cite{simp18} as a regular 
  counterpart of the \Scw\ space-tiime and its further generalizations \cite{franzin21, lobo20, kb22}   
  obtained from \RN\ and other singular \ssph\ metrics. The way of converting a metric singular at $r=0$
  to a regular one consists in replacing the spherical radius $r$ with the expression  $r(u) = \sqrt{x^2 + b^2}$, 
  thus eliminating the singularity and its close neighborhood from space-time and smoothly joining its
  remaining part to one more copy of this part.\footnote
		{Here, a new radial coordinate $x \in \R$ is used instead of $r$, since we keep the notation $r$ 
		for the spherical radius $r = \sqrt{-g_{\theta\theta}}$ to reconcile its geometric meaning with intuition, 
		Other radial coordinates are denoted by other letters to avoid confusion.}  
  In particular, the \Scw\ metric is converted to \cite{simp18}		
\beq            \label{SV}
			ds^2 = \bigg(1 - \frac{2M}{x^2 + a^2}\bigg)dt^2 
			- \bigg(1 - \frac{2M}{x^2 + a^2}\bigg)^{-1} dx^2 	- (x^2 + a^2) d\Omega^2.
\eeq   
  The resulting regular minimum of $r$ at $x=0$ is a \wh\ throat if it occurs in a static region (it happens
  if $a^2 > 4M^2$, it is an extremal \bh\ horizon if $a^2 = 4M^2$, and if $a^2 < 4M^2$, it is a bounce of 
  $r(x)$ as one of two cosmological scale factors in a Kantowski-Sachs anisotropic cosmology, 
  where $x$ is a temporal coordinate. This region is located between two simple horizons 
  $x = \pm \sqrt{4M^2 - a^2}$. Such a bounce inside a \bh\ is called a {\it black bounce} according to
  \cite{simp18}, and all space-times constructed in this way are often called ``black bounce space-times''
  or even simply ``black bounces.''  Numerous models found in this manner have gained much interest,  
  such \sph\ geometries were generalized to include  \cite{mazza21, xu21, shaikh21}, there followed 
  calculations of quasinormal modes, properties of gravitational waves and gravitational lensing in such 
  space-times, etc.  \cite{churilova19, yang21, guerrero21, tsukamoto21, islam21, cheng21, kb20,
  tsukamoto20, lima21, nascimento20, franzin22,vagnozzi22,pedrotti24,s_ghosh22,s_ghosh23}.
  This whole trend may be regarded as one of simple ways to simulate possible effects of quantum gravity 
  in the language of classical gravity, leaving aside any quantization details. The properties of new geometries 
  constructed in this way may also be of certain interest. One can also note that black bounces as the 
  phenomena of minimum $r(u)$ inside \bhs\ have been found as quite a common feature in a number 
  of solutions of GR and other theories in the presence of phantom scalar fields. Such space-times were 
  called ``black universes,'' their \bh\ interior at times after a bounce becomes an expanding universe 
  with isotropic  (in particular, de Sitter) late-time asymptotic, see, e.g.,
  \cite{bk-uni05, bk-uni06, bk-uni12, clem09, azreg11, trap18}.
  
  We can see that the basic feature of the black bounce concept is that a space-time singularity is cut away 
  together with its close environment, and the resulting regular space-time is almost doubled as compared 
  to the original one. There emerge new geometries and causal structures more complex than before, which 
  may cause additional interest but may be regarded as a shortcoming. Meanwhile, another phenomenological
  regularization method, tracing back to Bardeen's proposal, also deserves attention. It suggests, instead of 
  cutting out a piece of space-time containing a singularity, to modify the metric in such a way that it becomes 
  regular. 
  
  This approach was recently applied \cite{kb24} to a class of \ssph\ space-times whose Ricci tensors satisfy 
  the condition  $R^t_t = R^r_r$, which holds for a number of most important solutions of GR. As particular 
  examples, this paper considered the \Scw, \RN, and some Einstein-NED solutions with comparatively soft
  singularities at $r=0$. The present paper continues that study, trying to find regular counterparts of more 
  general (though not quite arbitrary)  \ssph\ space-times and consider the same examples that were 
  previously subject to SV-like modification in \cite{kb22}: Fisher's scalar-vacuum solution of GR with a
  massless minimally coupled scalar filed \cite{fisher48}, 
  and the simplest family of dilatonic \bh\ solutions of GR with interacting scalar and \emag\ fields 
  \cite{dil1, dil2, dil3, dil4}. Thus it becomes possible to compare different regular counterparts of the 
  same geometries that can be designated as SV-like (or black bounce) and Bardeen-like (or regular center)
  modifications, bearing in mind that some of them, or both, may simulate possible effects of quantum gravity.
  
  Of significant interest is the nature of possible sources of regularized space-times in the framework of GR
  or its alternatives. Thus, for a class of black-bounce space-times such sources were found in \cite{rahul22} 
  and \cite{canate22} within GR in the form of a combination of a minimally coupled phantom scalar field 
  with certain self-interaction potentials, and magnetic fields obeying some form of NED. It was noted that 
  neither NED alone nor a scalar field alone were able to produce a necessary source. A phantom field,
  violating all standard energy conditions, is quite a necessary source to account for a minimum of the spherical 
  radius $r$ that is present in all black-bounce metrics, while inclusion of NED makes it possible to adjust 
  the total stress-energy tensor (SET) $T\mN$ specified by a particular black-bounce metric. For SV-regularized 
  \Scw\ and \RN\ metrics, such sources were obtained explicitly in \cite{rahul22}. A similar method was used 
  for some cosmological space-times in \cite{kam22, kam24} .

  As shown in \cite{kb22}, a combination of a nonlinear \emag\ field and a minimally coupled self-interacting 
  scalar field provides a possible source for an {\it any\/} \ssph\ metric in the framework of GR. However, for 
  such an arbitrary metric, the scalar field must, in general, change its nature from one space-time region to 
  another, being canonical (that is, possessing positive kinetic energy) in some regions, and phantom (having 
  negative kinetic energy) in other ones, with smooth transitions between such regions. This kind of scalar fields,
  which are phantom in strong-field regions and canonical elsewhere, has been named ``trapped ghosts'' 
  by analogy with genies in bottles in oriental fairy tales \cite{trap10}. A number of globally regular 
  solutions have been found with such sources, including \whs\ and black universes, see, e.g., 
  \cite{kroger04, trap10, don11}. It was also argued that transitions between canonical and phantom regions 
  of such scalars might play a stabilizing role in space-times sourced by those scalars \cite{trap17, trap18}. 
  Explicit NED-scalar sources for SV-regularized Fisher and dilatonic \bh\ metrics were constructed in \cite{kb22}.
  
  In the case of Bardeen-like regularization, the spherical radius $r$ does not acquire a regular minimum,
  therefore, phantom matter as a source is not required. And indeed, according to \cite{kb24}, regularized 
  space-times with $R^t_t = R^r_r$ can be presented as magnetic Einstein-NED solutions without any scalar
  fields since this kind of regularization preserves the condition $R^t_t = R^r_r$. For more general space-times
  a scalar field source is required, but one may hope that this source will be only canonical. However, as we 
  will see, in the two examples under consideration, phantom regions of the source scalars necessarily emerge.     
  
  The paper is organized as follows. In Section 2, we make some preparations, recalling the regular 
  center conditions and the way to obtain a scalar-NED source for an arbitrary \ssph\ metric.
  Section 3 describes a regularization method applicable to sufficiently general space-time singularities 
  at $r=0$. In Section 4 we try to apply this method to the Fisher scalar-vacuum and 
  dilatonic \bh\ space-times. Section 5 contains some concluding remarks. The metric signature $(+\,-\,-\,-)$ 
  is adopted, and geometrized units are used, such that $8\pi G = c = 1$.

\section{Preliminaries} 
\subsection{Regularity conditions}

  In this subsection we recall some well-known facts to be used in what follows.
  Consider a pseudo-Riemannian space-time $\M$ with an arbitrary \ssph\ metric 
\beq 		\label{ds}
		ds^2 = A(x) dt^2 - \frac{dx^2}{A(x)} - r^2(x) d\Omega^2,\qq 
		d\Omega^2 = d\theta^2 + \sin^2\theta d\varphi^2,
\eeq  
  written here in terms of the so-called quasiglobal radial coordinate $x$ \cite{BR-book}. This choice of
  the radial coordinate is well suited for the description of any \ssph\ space-times including \bhs\ 
  (where horizons appear as regular zeros of $A(x)$ provided $r(x)$ is finite) and \whs\ (where throats 
  appear as regular minima of $r(x)$ provided $A(x) > 0$). 
  
  If somewhere in our space-time there is a location where $r\to 0$ under the condition $A > 0$, such a 
  location is called a center, and it is indeed a center of symmetry in spatial sections of $\M$. If $r\to 0$
  in a region where $A < 0$, it is a cosmological-type singularity instead of a center since $x$ can there 
  be used as a time coordinate. 
  
  A center is called regular if all algebraic curvature invariants are there finite and smooth, which includes, 
  in particular, the existence of a tangent flat space-time at this point. (We leave aside the behavior of 
  higher-order invariants involving derivatives of the Riemann tensor.) All these invariants are obtained as
  products and contractions of the four different nonzero Riemann tensor components
\bearr            \label{Riem}   
  		K_1 = - R^{01}{}_{01} = \Half A'', \qq   
  		K_2 =  - R^{02}{}_{02} =  - R^{03}{}_{03} = \Half \frac{A' r'}{r},
\nnn
		K_3 = - R^{12}{}_{12} =  - R^{13}{}_{13} = A\frac{r''}{r} + \Half \frac{A' r'}{r},\qq
		K_4 = - R^{23}{}_{23} = \frac 1{r^2} (A r'{}^2 -1),   
\ear 
  where the coordinates are numbered as $(0,1,2,3) = (t,x,\theta,\varphi)$, and the
  prime denotes $d/dx$. Since the Riemann tensor $R\MN{}\mn$ is pairwise diagonal,
  the Kretschmann scalar $\cK = R_{\mu\nu\rho\sigma}R^{\mu\nu\rho\sigma}$ is a sum of squares,
  $\cK = 4K_1^2 + 8K_2^2 + 8K_3^2 + 4K_4^2$, hence for its finiteness it is necessary and sufficient that
  each $K_i$ is finite, and if so, it guarantees that all algebraic invariants of the Riemann tensor are finite. 
  
  For the metric \rf{ds} this requirement implies (note especially the expression of $K_4$) that if 
  $r(x) \to 0$ at some $x \to x_0$, then
\beq             \label{reg-c}
			A(x) = A_0 + \cO(r^2), \qq  A(x) r'{}^2(x) = 1 + \cO(r^2), \qq  A_0 = \const >0,
\eeq   
  where the symbol $\cO(r^2)$ means a quantity of the same order as $r^2$ or smaller. 
  
  In the important case considered in \cite{kb24}, where the Ricci tensor satisfies the condition 
  $R^0_0 - R^1_1 =0$, the corresponding component of the Einstein equations
\beq 			\label{EE}
		G\mN \equiv R\mN - \Half \delta\mN R = - T\mN 
\eeq  
  leads to the condition $r''(x) =0$, and almost without loss of generality we can put $r(x) \equiv x$
  (rejecting only ``flux tubes'' with $r= \const$), so that, after an evident rescaling of $t$ and $x$, 
  the quasiglobal coordinate coincides with the more frequently used \Scw\ coordinate $r$.  
  Since now $r' \equiv 1$, in the regularity conditions \rf{reg-c} we must put $A_0=1$, while the
  ${0 \choose 0}$ component of \eqs \rf{EE} can be rewritten in the integral form as
\beq           \label{A(r)}
		A(r) = 1 - \frac 1 {r} \int T^0_0\, r^2 dr.
\eeq  
  Considering here integration from $r=0$, we see that the regularity condition $A(r) = 1 + \cO(r^2)$
  as $r\to 0$ holds as long as the density $\rho = T^0_0$ is finite at $r=0$.  
  
\subsection{Scalar--NED sources for spherical space-times}

  As shown in \cite{kb22}, {\it any\/} (sufficiently smooth) metric \rf{ds} is a solution to the 
  Einstein equations with a source given by a non-interacting combination of a scalar  field $\phi$ with a 
  certain self-interaction potential $V(\phi)$ and a nonlinear \emag\ field with some Lagrangian 
  density $\cL(\cF)$, where $\cF = F\mn F\MN$, and $F\mn = \D_\mu A_\nu - \D_\nu A_\mu$ is
  the \emag\ field tensor. Let us briefly reproduce this result showing how to obtain the functions 
  $V(\phi)$ and $\cL(\cF)$ from given $A(x)$ and $r(x)$.
  
  Consider the action $S_m$ of a combination of a scalar field $\phi(x)$ and NED minimally coupled 
  to gravity, 
\beq                       \label{S_m}
		S_m = \int \sqrt{-g}\, d^4x \big[ 2 h(\phi) g\MN \D_\mu\phi \D_\nu\phi - 2V(\phi) -\cL(\cF) \big],
\eeq
  where $h(\phi)$ is a function related to the parametrization freedom of $\phi$; if $h(\phi) > 0$, 
  we can absorb it by redefining $\phi$ as $d\phi_{\rm new} = \sqrt{h(\phi)} d\phi$, so that 
  $h(\phi_{\rm new}) \equiv 1$, and we are thus dealing with a canonical scalar field having positive 
  kinetic energy. If $h(\phi) < 0$, a similar reparametrization leads to $h(\phi_{\rm new}) \equiv -1$ 
  that corresponds to a phantom scalar field. If $h(\phi)$ somewhere changes its sign, then the scalar $\phi$ 
  changes its nature from canonical to phantom, as happens in ``trapped ghost'' models of \whs\ where 
  the scalar turns out to be canonical in a weak field region and phantom in a strong field one \cite{trap10}.  

  With such matter,  the SET in the Einstein equations \rf{EE} has the form 
  $T\mN = T\mN[\phi] + T\mN[F]$, where
\bearr                   \label{SET-phi}
			T\mN[\phi] = 2 h(\phi) \D_{\mu}\phi\D^{\nu}\phi 
				- \delta\mN \big[ h(\phi) g^{\rho\sigma}\D_\rho \phi \D_\sigma\phi -V(\phi) \big],
\yyy                    \label{SET-F}
			T\mN[F] = - 2 \mathcal{L_F} F_{\mu\sigma} F^{\nu\sigma} 
					+\frac 12 \delta\mN \mathcal{L(F)},
\ear 
  with $\mathcal{L_F} = d\cL/d\cF$. The scalar and \emag\ field equations that follow from \rf{S_m} are
\bearr         \label{eq-phi}
			2 h(\phi) \nabla_{\mu}\nabla^{\mu}\phi 
					+ \frac{d h}{d\phi} \phi^{,\mu}\phi_{,\mu} + \frac{d V(\phi)}{d \phi} =0,
\yyy		\label{eq-F}
			\nabla_\mu(\mathcal{L_F}F\MN) = 0.
\ear

  In accord with the symmetry of $\M$, we assume $\phi = \phi(x)$ and suppose the existence of only 
  a radial magnetic field, so that the only nonzero components of $F\mn$ are 
  $F_{\theta\varphi} =-F_{\varphi\theta} = q \sin\theta$, where $q$ is a monopole magnetic charge. 
  Then \eq \rf{eq-F} is trivially satisfied, while the invariant $\cF$ is expressed as
  $\mathcal{F} = 2 q^2/r^4$, independently from the choice of $\cL(\cF)$. The SETs \rf{SET-phi} 
  and \rf{SET-F} then take the form
\bearr         \label{T-phi}
		T\mN[\phi] = h(\phi) A(x) \phi'{}^2 \diag (1, -1, 1, 1) + \delta \mN V(\phi), 
\yyy           \label{T-F}		
		T\mN[F] = \frac 12 \diag\Big(\cL,\ \cL,\ \cL - \frac{4q^2}{r^4} \cL_\cF,\
				\cL - \frac{4q^2}{r^4} \cL_\cF\Big). 
\ear  

  The scalar field equation and the nonzero components of the Einstein equations for the metric \rf{ds} read
\bearr                  \label{phi''}           
		\frac{2h}{r^2}\Big(A r^2\phi'\Big)' + A \phi'{}^2 \frac{dh}{d\phi} = \frac{dV}{d\phi} ,
\\ \lal                 \label{EE0}
  	 G^t_t = \frac{1}{r^2} [-1 + A (2 r r'' + r'{}^2) + A' r r']  = - T^t_t,,
\\ \lal  	        \label{EE1}
	 G^x_x = \frac{1}{r^2} [-1 +  A' r r' + A r'{}^2] = -T^x_x,
\\ \lal           \label{EE2}
	 G^\theta_\theta = G^\varphi_\varphi = \frac{1}{2r} [2 A r'' + r A'' + 2 A' r'] = - T^\theta_\theta.
\ear   
  Two differences of these equations, \rf{EE0}-\rf{EE1} and \rf{EE0}-\rf{EE2}, give, respectively,
\bearr             \label{01}
		2 h (\phi) \phi'^2  = - 2 \frac{r''}{r}, 
\\ \lal  			\label{02}
		\frac {2q^2}{r^4} \cL_{\cF} = \frac 1{2 r^2}\Big[2- A (r^2)'' + A'' r^2 \Big].
\ear
  
  There can be different ways of finding the material quantities from given $A(x)$ and $r(x)$.
  Maybe the simplest one is as follows. From \eqn{01} we know the quantity $h(\phi)\phi'{}^2$ 
  as a function of $x$. Using the parametrization freedom of $\phi$, we can choose $\phi(x)$ as any 
  convenient monotonic function, after which the function $h(\phi)$ will be known, and it is important 
  that its sign depends only on the sign of $r''(x)$ but not on the choice of $\phi(x)$. More precisely, 
  the scalar is canonical if $r'' < 0$, it is  phantom if $r'' >0$,  and in the general case it can change 
  its nature from one region to another. Next, with known $\phi(x)$ and $h(x)$, from \eqn{phi''}
  we can find by integration the potential $V$ as a function of $x$ or $\phi$, choosing the integration
  constant so that $V=0$ at spatial infinity. Then, for example, \eqn{EE1} makes it possible to 
  find $\cL(x)$ as a function of $x$: 
\beq           \label{L(x)}
  		\cL(x) + 2V = - 2 A \frac{r''}{r} + \frac{2}{r^2} (1 - A' r r' - A r'{}^2), 
\eeq  
  and since $\cF(x) = 2q^2/r^4(x)$ is known, the functions $\cL(\cF)$ is also found. It remains 
  to verify the correctness of calculations by making sure that $\cL' = \cL_{\cF} \cF'$, with 
  $\cL_{\cF}$ determined from \eqn{02}. 
  
  An alternative way is to determine $\cL(x)$ by integration from \eqn{02}, also using the fact that 
  $\cF(x) = 2q^2/r^4(x)$ is known. After that, $V(x)$ can be found from \eqn{L(x)}, and with known 
  $\cF(x)$ and $\phi(x)$, it is straightforward to obtain $V(\phi)$ and $\cL(\cF)$. In this case, 
  the scalar equation \rf{phi''} is left aside, and this equation is known to follow from \rf{EE0}--\rf{EE2}.
  
  An observation of interest is that the field equations only imply that the sum $\cL(\cF) + 2V(\phi)$
  must tend to a finite limit at the regular center $r=0$, but it follows from nowhere that this is true for 
  each of these quantities taken separately. In particular, from \eqn{02} it follows that 
  $\cF \cL_\cF$ is finite at a regular center, where $\cF \sim r^{-4} \to \infty$, but then $\cL$ is 
  determined by the integral $\int  \cL_\cF d\cF$, which is, in general, logarithmically divergent.
  A similar conclusion is achieved for the potential $V(\phi)$ after an analysis of \eqn{phi''}.
  
  The important special case where $r'' \equiv 0$, corresponding to the condition $T^t_t = T^x_x$
  according to \eqn{01}, has been considered in \cite{kb24}. In this case, putting $r \equiv x$, 
  we can choose a nonlinear magnetic field obeying NED as the only source of gravity. The 
  corresponding Bardeen-like regularization was obtained in \cite{kb24} for \Scw, \RN\ and 
  some singular Einstein-NED solutions. Thus, for the \Scw\ solution, one naturally reproduces 
  Bardeen's original metric \rf{bardeen} with a de Sitter asymptotic behavior at $r=0$, but other
  versions of such regularization were also indicated, leading to an asymptotically Minkowski regular 
  center characterized by zero curvature. A similar choice of central asymptotic behaviors was shown 
  to be possible for \RN\ and other singular solutions obeying the condition $R^t_t = R^x_x$,
  and magnetic NED sources for their regularized versions were indicated.     
  
  It was also noted that electric instead of magnetic field sources are also possible, but then, in 
  the case of metrics with a regular center, we would inevitably obtain NED theories with different
  Lagrangians $\cL(\cF)$ acting in different parts of space, as shown in the studies devoted 
  to \sph\ NED-Einstein configurations \cite{kb01-NED, kb01-comm}. Curiously, the same 
  phenomenon was observed in electric sources of black-bounce space-times considered in 
  \cite{we24}, for which it is not inevitable but still happens in a number of examples. In the present 
  study we are again dealing with systems possessing a regular center, and therefore, to avoid this
  kind of ambiguity in $\cL(\cF)$, we restrict ourselves to magnetic NED solutions.
  
\section{Creating a regular center}

  A black-bounce-like replacement was suggested in \cite{kb22} for sufficiently general \ssph\
  space-times with a singularity at $r = 0$, not necessarily with $R^t_t \ne R^x_x$, so that 
  $r(x) \not\equiv x$), whose metrics are better presented in radial coordinates other than $r$. 
  Two examples were considered there: Fisher's solution of GR with a canonical scalar field as a source 
  \cite{fisher48}, and a dilatonic \bh\ \cite{dil1, dil2, dil3, dil4}. Let us now consider the same 
  space-times and try to cure the singularity by making $r=0$ a regular center satisfying the 
  conditions \rf{reg-c}.
  
  Thus we are dealing with the general \ssph\ metric in the form \rf{ds}. It is hard to suggest a universal 
  recipe of singularity smoothing for any behavior of the metric functions, therefore, let us suppose that 
  near a singularity that occurs at $x=0$, the functions $A(x)$ and $r(x)$ can be presented in the form
\bearr                 \label{A-sing}
			A(x) = x^a A_1(x), \qq  a = \const \ne 0, \qq A_1 (0) \ne 0,
\yyy	                 \label{r-sing}
			r(x) = x^b r_1(x), \qq \ \ b  = \const > 0, \qq\  r_1(0) \ne 0,
\ear  
  where $A_1(x)$ and $r_1(x)$ are some sufficiently smooth functions.
  
  The desired regularization must smooth out the metric at $r=0$ but leave it (almost) invariable
  far from it. Since now we have to modify two metric functions, this procedure requires two steps.
  
\paragraph {Step 1.} To begin with, a universal recipe for making $A(x)$ regular is to make finite 
  the matter density $\rho$ at $r=0$, as was done in \cite{kb24}. Then \eqn{A(r)} gives the desired
  solution in terms of the coordinate $r$. However, in other coordinates this relation changes its form,
  for exmple, in terms of the coordinate $x$ used here it reads
\beq                     \label{A(x)}
			A(x) = \frac{1}{r'^2}\bigg[1 - \frac 1{r(x)} \int \rho(x) r^2(x) r'(x) dx  \bigg],
\eeq   
  and without knowledge of $r(x)$ it does not directly give a solution. 
  and regularity at $x=0$ also implies a finite value of $r'(0)$. Therefore, let us try other ways to cure 
  the metric function $A(x)$.
  
  First, assuming $A_1(0) > 0$, the function $A(x)$ is easily regularized by a replacement similar 
  to the one used by SV \cite{simp18},
\beq 			\label{reg-x}
                 x \mapsto  u(x) = \sqrt{x^2 + c^2}, \quad  c = \const >0; 
\qq
			A(x) \mapsto \Areg(x)  = A(u)
						= c^a \Big(1 + \frac {x^2}{c^2}\Big)^{a/2} A_1(u),
\eeq  
  where the replacement $A_1(x) \mapsto A_1(u)$ is needed to avoid a nonzero derivative $A'_1(0)$

  If $A_1(0) < 0$, which happens if the singularity is located inside a \bh, our task is to make $A(0)$ 
  not only finite but also positive. Therefore, let us assume, instead of \rf{A-sing},
   a behavior of $A(x)$ like that considered in \cite{kb24},
\beq                   \label{A-sing-BH}
		A(x) = (1 - B/x^m) A_1(x), \qq B=\const >0,\qq  m = -a = \const \geq 0,
\eeq     
  with $A_1(x)$ as in \rf{A-sing}. If $m > 0$, this $A(x)$ acquires the form we need after the
  Bardeen-like replacement in the expression $(1 - B/x^m)$
\beq            \label{x-reg}
			x^{-m} \ \mapsto \  \frac {x^{2mn}}{u^{2mn + m}},
\qq
			n = \const \geq 1/m,   \qq   c = \const > 0, \qq  u = \sqrt{x^2 + c^2},
\eeq
  and, as before, replacing $A_1(x) \mapsto A_1(u)$. For example, with $m=n=1$ the procedure 
  \rf{x-reg} precisely reproduces Bardeen's (with $x$ instead of $r$). More generally, our replacements
  provide a regular behavior of $A(x)$ at $x=0$, preserving its original properties at sufficiently large $x$. 
  
  However, in the case $m=0$, such that $A$ tends to a finite negative value, this method does 
  not work, and it remains to seek how to regularize $A(x)$ in each particular case ``individually''.

\paragraph{Step 2.}
   If we simply change $x \mapsto u= \sqrt{x^2 + c^2}$, the center will be cut away along with 
   its neighborhood as in the black bounce paradigm. Instead, let us try the following replacement:
\beq
			r (x) \mapsto K x (1 + \alpha x^2)^N r_1(u), \qq  
					K, \alpha, N = \const.
\eeq   
  The constant $K$ must be found in such a way as to satisfy the second regularity condition \rf{reg-c}, 
  which now reads $A r'{}^2 = 1 + \cO(x^2)$ near $x=0$, but this $\cO(x^2)$ is only provided if we 
  replace, in addition, $x \mapsto u= \sqrt{x^2 + c^2}$ in the argument of $r_1(x)$ to prevent the
  emergence of a term linear in $x$. Thus we have 
\beq
			K = \frac {1}{r_1(c) \sqrt{c^a A_1(c)}}
\eeq   
  Now the conditions  \rf{reg-c} are satisfied  completely. 
   
  It remains to provide the correct behavior of $r(x)$ at large $x$, the same as in the original metric,
  and this can be achieved by a proper choice of the constants $N$ and $\alpha$:
\beq 			\label{reg-r}
		N = \frac{b-1}{2}, \qq  \alpha = K^{-1/N}= \Big(r_1(c) \sqrt{A_1(c) c^a}\Big)^{2/(b-1)}.
\eeq       

  This scheme does not work in the case $b =1$, when we must have $r \approx Kx$ at small $x$
  but $r_0 = r_1(0) \ne K$. If $r = r_0 x$ holds at all $x$, then, making a rescaling of the coordinates $x$ 
  and $t$, we arrive at the simpler situation described in \cite{kb24} with $r \equiv x$. Otherwise, we can 
  use the replacement $r_0 \mapsto K \then r \approx Kx$ at small $x$, which is necessary for \rf{reg-c},
  but then we have to modify $r(x)$ in such a way as to restore the factor $r_0 x$ at larger $x$. This 
  can be achieved by replacing (in the relevant factor only) in the original $r(x)$,
\beq
			r_0 x \mapsto x s(x), \qq \text{where} \qq s(x) \to K \ (x\to 0), \ \ \ s(x) \to r_0\ (x\to \infty).
\eeq   
  For example, one may take $s(x) = K + (r_0 - K) \tanh^2 x$. This completes the description of a 
  regularization procedure under the assumptions \rf{A-sing} and \rf{r-sing}.  
  
\section{Two examples} 
\subsection{Fisher's solution}   

  Let us apply the described procedure, as the first example, to the solution of GR with a canonical massless 
  scalar field $\Phi$, discovered by I.Z. Fisher in 1948 \cite{fisher48} and later repeatedly rediscovered 
  \cite{JNW68, Wyman81}
  
  This solution corresponds to the source \rf{S_m} with $h(\phi) \equiv 1$, $V \equiv 0$, without an 
  \emag\ field, and can be written as (see, e.g, \cite{BR-book})  		
\bearr             \label{fish}
		ds^2 = \Big(1- \frac{2k}{y}\Big)^a dt^2 - \Big(1- \frac{2k}{y}\Big)^{-a} dx^2 
					- y^2 \Big(1- \frac{2k}{y}\Big)^{1-a} d\Omega^2,
\nnn
	    \Phi = \pm \frac{\sqrt{1-a^2}}{2} \ln \Big(1- \frac{2k}{y}\Big), 
\ear    
  where $k > 0$ and $a \in (-1, 1)$ are integration constants, such that $m = ak$ has the meaning of the 
  \Scw\ mass, and the quantity $\pm k\sqrt{1-a^2}$ has the meaning of a scalar charge. The coordinate 
  $y$ ranges from $2k$ to infinity, the value $y=2k$ being a naked singularity with $r(y)=0$. So, let us 
  rewrite the metric \rf{fish} in terms of $x = y - 2k$, an accord with the notations used: 
\beq             \label{fish-x}
		ds^2 = \bigg(\frac{x}{x+2k}\bigg)^a dt^2 - \bigg(\frac{x}{x+2k}\bigg)^{-a} dx^2 
					- x^{1-a} (x+2k)^{1+a} d\Omega^2,
\eeq   
  and we seek the regularized metric
\beq              \label{ds-reg}
		ds^2= \Areg(x) dt^2 -\frac {dx^2}{\Areg(x)} - \rreg^2(x) d\Omega^2.
\eeq  
  Applying the first step \rf{reg-x} to $A(x) = x^a/(x+2k)^a$, where the power 
  $a$ has the same meaning in \rf{reg-x} and \rf{fish-x}, we obtain the regularized $A(x)$ as
\beq             \label{Areg-fish}
		\Areg(x) = (x^2 + c^2)^{a/2} \big(\sqrt{x^2 + c^2} + 2k\big)^{-a}.
\eeq  
  Note that here $A_1(x) = (x+2k)^{-a}$. The resulting regular nature of $\Areg(x)$ is illustrated in 
  Fig.\,1(a).
   
  At the second step, we have $r_1(x) = (x+2k)^{1+a}$, and we obtain the constants and the regularized 
  function $r^2(x)$ as
\bearr                  \label{rreg-fish}
  		b = \frac{1-a}{2}, \qq N = -\frac{1+a}{4},  \qq
  		K = \frac {c^{-a/2}}{\sqrt{c+2k}},  \qq \alpha = c^{\tfrac{2a}{1+a}}  (c+2k)^{-\tfrac{2}{1+a}},
\nnn
		\rreg^2(x) = \frac{c^{-a} x^2}{c + 2k} 
		\big(1 + \alpha x^2\big)^{-(1+a)/2}\Big(\sqrt{x^2 + c^2} + 2k\Big)^{1+a}. 
\ear
  The behavior of the original and regularized functions $A(x)$ and $r(x)$ is shown in Fig.\,1 for some
  values of the parameters. The values of $c$, which must be quite small if we try to refer to quantum
  gravity effects, are here chosen to be rather large for illustration purposes.
\begin{figure*} \centering   
\includegraphics[scale=0.3]{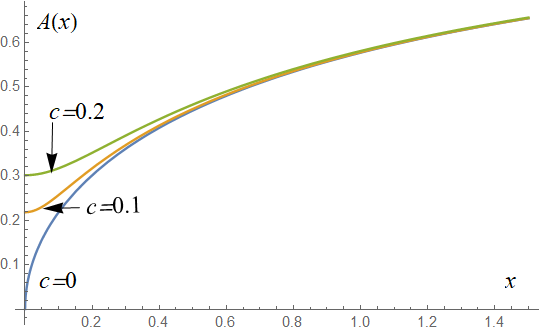}   
\includegraphics[scale=0.42]{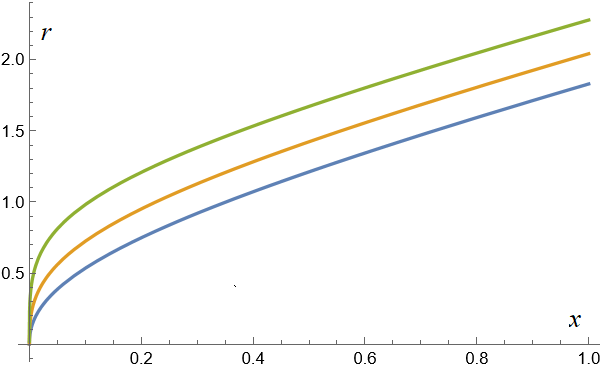}
\includegraphics[scale=0.42]{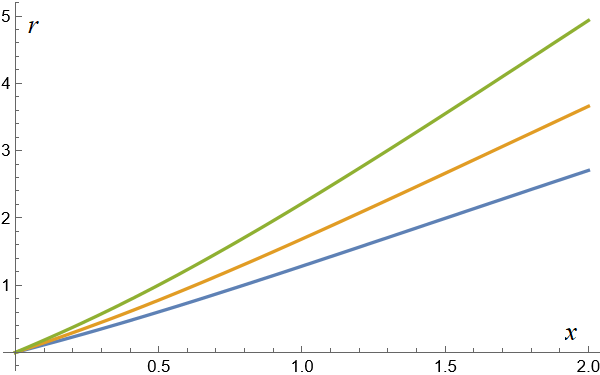} \\  
            (a)   \hspace{2in}   (b)   \hspace{2in}   (c) 
\caption{\protect\small 
		Fisher's solution and its regularization. (a) The function $A(x)$ for Fisher's solution ($c=0$) 
		with $k =1, a = 1/2$, and its regularized version with two values of $c$.
		(b) The spherical radius $r(x)$ in Fisher's solution with $k=1$ and $a = 0.1, 0.3, 0.5$ (bottom-up).
		(c) The spherical radius $r(x)$ in regularized Fisher's solution with $k=1$, $c=0.2$, 
		and $a = 0.1, 0.3, 0.5$ (bottom-up).
		}   \label{fig-fish1}
\end{figure*}                                           

  Now, if we try to find a possible source for this metric with the action \rf{S_m}, it turns out that any of 
  the two ways described in Sec.\,2, either with \eqn{phi''} or with \eqn{02}, leads to very complex  
  integrals that evidently cannot be calculated analytically. 
  For example, the expression of $\cL'(x) = d\cL(\cF(x))/dx$ obtained from \eqn{02} reads
  (notations: $u = \sqrt{x^2+c^2}, \ c_1 = c^{2a/(1+a)},\ c_2 = (c+2k)^{2/(1+a)}$)
\bearr  \nqq
  \frac{d\cL(x)}{dx} =
	\frac {c^a (2 k+u)^{-a-2} } {x^3 u \left(c_1 x^2+c_2\right)^{(1-a)/2} (c+2k)} 
	\Big(2c_1x^4 + [(1-a) c_1(c^2 + 2ku) + c_2 (3+a) ] x^2	+2 c_2 (2 k u+ c^2)\Big)
\nnn  \times 
	\Bigg(\frac {c^{-a} u^a }{(c_1 x^2 + c_2)^{(1+a)/2}(c+2k)}
   \bigg[\frac{(a+1) (a+3) x^4 c_1^2 (2 k+u)}{\left(c_1 x^2+c_2\right)^2}-\frac{5 (a+1)
   c_1 x^2 (2 k+u)}{c_1 x^2+c_2}
\nnn
	-\frac{2 (a+1)^2 c_1 x^4}{u \left(c_1 x^2+c_2\right)}+\frac{a (a+1)
   x^4}{u^2 (2 k+u)}-\frac{(a+1) x^4}{u^3}+\frac{5 (a+1) x^2}{u}+2 (2 k+u)\bigg]
\nnn
	-\frac{2 a k x^2  c^{-a} u^{a-5} \left[2 (a-1) k u x^2+c^4-c^2 \left(x^2-2 k u\right)-2 x^4\right]}
		{(2 k+u) (c_1 x^2 + c_2)^{(1+a)/2}(c+2k)}  - 2 (c+2 k)\Bigg).
\ear
  However, the behavior of the function $\cL(\cF)$ turns out to be rather simple, as illustrated by the plots 
  of $\cL(x)$ in Fig.\,2(a) and for $\cL(\cF)$ in Fig.\,2(b). As was expected, there is a logarithmic divergence 
  as $\rreg(x) \to 0$ ($\cF \to \infty$)  and a sufficiently rapid decay as $\rreg(x) \to \infty$ ($\cF \to 0$),
  without a correct Maxwell limit at small $\cF$ which would require $\cL \approx \cF$.

\begin{figure*} \centering   
\includegraphics[scale=0.45]{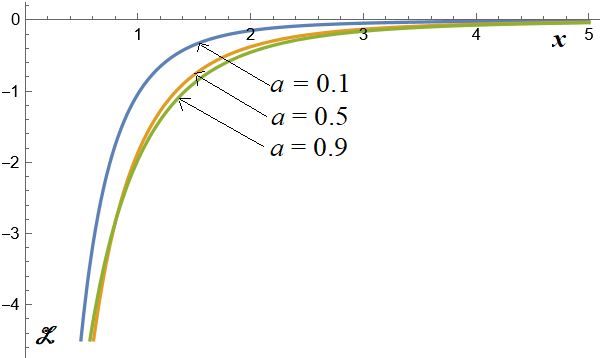} \qq        
\includegraphics[scale=0.45]{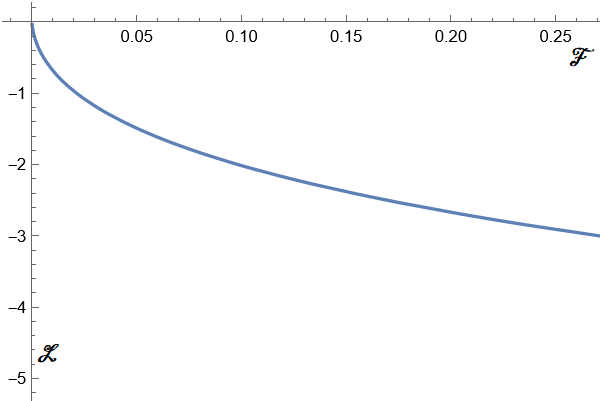}\\
              (a)   \hspace{2in}   (b)
\caption{\protect\small 
		A possible source for the regularized Fisher solution: 
		(a) The function $\cL(x)$ for $k=1,\ c = 0.2$, and $a = 0.1, 0.5, 0.9$.
		(b) The function $\cL(\cF)$ for $k=1,\ c=0.2,\ a = 0.5$, and $q = 1$ (so that $\cF = 2/\rreg^4$).
		}  \label{fig-fish2}      
\end{figure*}                                           

  The scalar source $\phi$ of the regularized solution (not to be confused with $\Phi$ from the original 
  singular solution) is characterized by the potential $V(\phi(x))$ and the function $h(\phi(x))$ whose 
  sign is of particular interest for us since it determines the canonical or phantom nature of the source, 
  whereas the function $\phi(x)$ may be chosen arbitrarily, it should only be monotonic. 
  Assuming for convenience 
\beq                  \label{phi}
		\phi = \arctan{x/c}, 
\eeq
  with the functions $A=\Areg(x)$ and $r= \rreg(x)$ according to \rf{Areg-fish} and 
  \rf{rreg-fish}, we find from \eqn{01}
\bearr   \nq
			h(\phi(x)) = -\frac{r''}{r} \frac{(x^2+c^2)^2}{c^2} 
			= - \frac{(1 + a) u}{4 c^2 (2 k + u)^2 [(c + 2 k)^{2/(1 + a)} + c^{2 a/(1 + a)} x^2]^2}
\nnn \times			
			\Big[ 6 c^2 (c + 2 k)^{4/(1 + a)} (2 k + u) 
			-  6 c^{4 + 2 a/(1 + a)} (c + 2 k)^{2/(1 + a)} (4 k + u) - (1- a) c^{4 + 4 a/(1 + a)} (4 k + u) x^2 
\nnn			
			+ (c + 2 k)^{4/(1 + a)} [8 k + (3 + a) u] x^2 - 4 (1 - a) c^{4 a/(1 + a)} k^2 u x^4 
\nnnv			
			-  2 c^{2 a/(1 + a)} (c + 2 k)^{2/(1 + a)} x^2 \{12 k^2 u (1+c^2) + 2 x^2 [(3 + a) k  + u ]
							             +  c^2  x^2 [2 (7 + a) k + (1 + a) u ]\}
\nnnv			
			 +  2 c^{2 + 4 a/(1 + a)}  x^2 [u (-2 k^2 + x^2) + 2 a k (k u + x^2)]\Big],  
\ear  
  where, as before, $u = \sqrt{x^2 +c^2}$. As can be seen from Fig.\,3(a), we have obtained $h < 0$ at 
  small values of $x$, near the new regular center. At larger $x$, the scalar is canonical, and thus we are 
  dealing with what has been called a trapped ghost scalar, showing a phantom nature only in a strong field 
  region \cite{trap10}. It can also be noticed for the curves in Fig.\,1(b) that $r'' < 0$, corresponding to 
  $h >0$, while Fig.\,1(c) shows curves with small positive $r''$ (at least cllose to $x=0$), corresponding 
  to $h <0$. The behavior of the potential $V(\phi(x))$, determined using \eqn{L(x)}, is shown in Fig.\,3(b); 
  its $\phi$ dependence is easily reproduced according to \rf{phi} by substituting $x = c \tan \phi$.
\begin{figure*} \centering   
\includegraphics[scale=0.55]{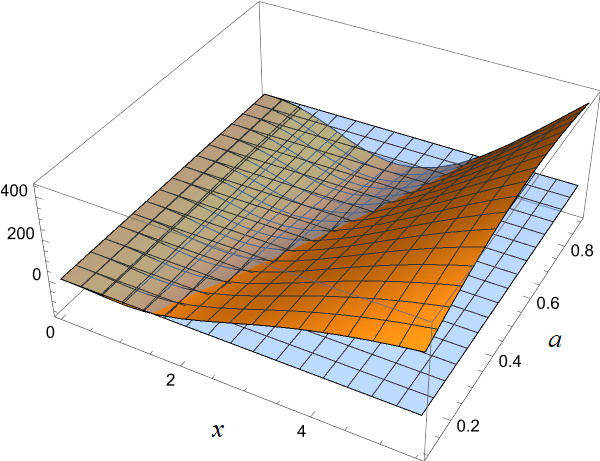}  \qq
\includegraphics[scale=0.6]{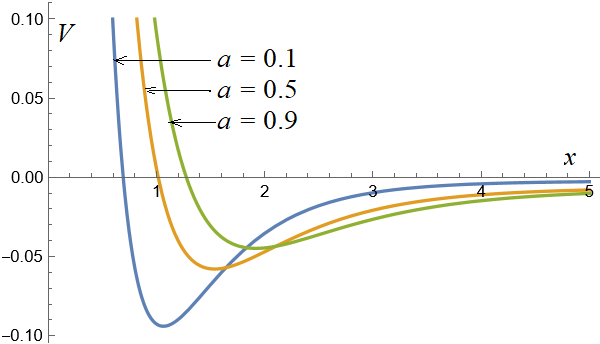}\\
              (a)   \hspace{2in}   (b)
\caption{\protect\small 
		A possible source for the regularized Fisher solution: 
		(a) The function $h(\phi(x))$ at $k=1,\ c=0.1$, and different values of $a$. 
		The plane $h =0$ is drawn to visualize the region with $h < 0$.
		(b) The function $V(\phi(x))$ for $k=1,\ c=0.2,\ a = 0.1, 0.5, 0.9$. 
		}  \label{fig-fish3}      
\end{figure*}                                           
    
\subsection{A dilatonic black hole}

  Dilatonic \bhs\ are described by special solutions of GR with a source consisting of a massless scalar field 
  $\Phi$ interacting with an electromagnetic field according to the action 
\beq              \label{S-dil}
		S_{\rm dil} = \Half \int \sqrt{-g}\, d^4x \big[2 g\MN \Phi_{,\mu}\Phi_{,\nu}
						- \e^{2\lambda\Phi} F\mn F\MN \big],									 
\eeq  
  where $\lambda$ is a coupling constant. The corresponding \bh\ solution can be written in terms of the metric 
  \rf{ds} with \cite{dil1, dil2, dil3, dil4}
\beq                                       \label{ds-dil}
		A(x) = \Big(1 - \frac{2k}{x}\Big)\Big(1 + \frac px \Big)^{-2/(1+\lambda^2)},
		\qq
				r^2(x) = x^2 \Big(1 + \frac px \Big)^{2/(1+\lambda^2)},
\eeq
  while the scalar $\Phi$ and the electric field $\vec E$ are given by
\beq               \label{phi-dil}
			\Phi = - \frac {\lambda}{1+\lambda^2} \ln \Big(1+ \frac px \Big),
			\qq
			2{\vec E}{}^2 = -F\mn F\MN = \frac {Q^2}{r^4(x)} \e^{-4\lambda \Phi},  
\eeq  
  where $Q$ is the electric charge, $k$ is one more integration constant, and
  $p = \sqrt{k^2 + Q^2(1+\lambda^2)} - k > 0$. 

  Consider the case $\lambda = 1$ related to string theory 
{dil2, dil3}. Then the metric has the simple form
\beq                  \label{ds-dil1}
		ds^2 = \frac{1-2k/x}{1+p/x} dt^2 - \frac{1+p/x}{1 -2k/x} dx^2 - x (x+p) d\Omega^2.
\eeq
  with a horizon at $x = 2k$, a singularity at $x =0$, and the \Scw\ mass $m = k + p/2 = Q^2/p$. 
  
  The singularity $x=0$ is located beyond the horizon, and since there $A \to -2k/p =\const$, it seems
  to be the relatively complex case, however, fortunately, due to the simple structure of $A(x)$, it is 
  readily regularized by the substitution \rf{x-reg} with $m=1$. Taking there $n=1$, we obtain
\beq             \label{Areg-D}
			\Areg(x) = \frac {(x^2+c^2)^{3/2} - 2kx^2}{(x^2+c^2)^{3/2} + p x^2}.
\eeq  
   For Step 2 we have $r_1(x) = \sqrt{x + p}$, consequently, 
\bearr             \label{rreg-D}
			a =0, \qq b = 1/2, \qq N = -1/4, \qq \alpha = 1/(c+ p)^2,
\nnn
			\rreg^2 (x) = \frac{x^2 \big(\sqrt{x^2 + c^2} + p\big)} {\sqrt{x^2 + (c+p)^2}},
\ear
  and we can write the regularized metric in the form \rf{ds-reg} with 
  $\Areg$ and $\rreg$ given by \rf{Areg-D} and \rf{rreg-D}. The smoothed form $\Areg$ of $A(x)$ is 
  shown in Fig.\,4 for particular values of the parameters. As to the radius \rf{rreg-D}, its $x$ dependence
  is very close to $\rreg \approx x $ and weakly depends on the parameters $p$ and $c$, so it does 
  not make sense to show its plots.
\begin{figure} \centering   
\includegraphics[scale=0.4]{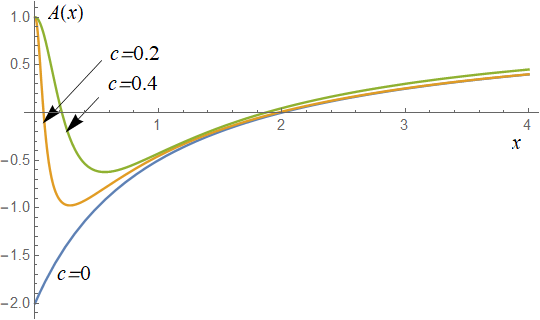}   
\caption{\protect\small 
		The dilatonic \bh\ solution \rf{ds-dil1} and its regulrization.
		 The function $A(x)$ for with $k = p =1$ ($c=0$) and its regularized versions with two 
		 nonzero values of $c$.
		}
\end{figure}                                           
  
  As in the previous subsection, let us use \eqn{02} to obtain an expression for $d\cL/dx$ for a possible
  NED source of the regularized metric (as before, $u = \sqrt{x^2 + c^2}$):
\bearr  \nqq
		\frac{d\cL}{dx} = 
	-\frac{2 \Big[(x^2/u) \left((c+p)^2+x^2\right) + (p+u) \left(2(c+p)^2+x^2\right) \Big]}
		{x^3 (p+u)^2 {u^3 \left((c+p)^2+x^2\right)^3 \left(p x^2+u^3\right)}}
\nnn \times
	\bigg\{-\left(u^3-2 k x^2\right) \Big[-2 u^3 (c+p)^4 (p+u)+u^2 x^2 (c+p)^2 
		\Big(u(p+u)\left((c+p)^2+1\right) - 5 (c+p)^2\Big)
\nnn 
	+x^6 \left(2 (c+p)^2+p u^3+u^4-3 u^2\right)
		+x^4 (c+p)^2 \left((c+p)^2+2 p u^3+2 u^4 - 8 u^2\right)+x^8\Big]
\nnn
   		 -2{u^3 \left((c+p)^2+x^2\right)^{5/2} \left(p x^2+u^3\right)}\bigg\}.
\ear
  Unlike the case with Fisher's solution, it is now possible to integrate this expression analytically,
  but the result is too cumbersome to handle. Instead, in Fig.\,5 we present examples of $\cL(x)$ and 
  $\cL(\cF)$ for some values of the parameters, obtained by numerical  integration. It is observed, 
  in particular, that in this case the NED Lagrangian exhibits a Maxwell-like behavior at small $\cF$.
\begin{figure*} \centering   
\includegraphics[scale=0.55]{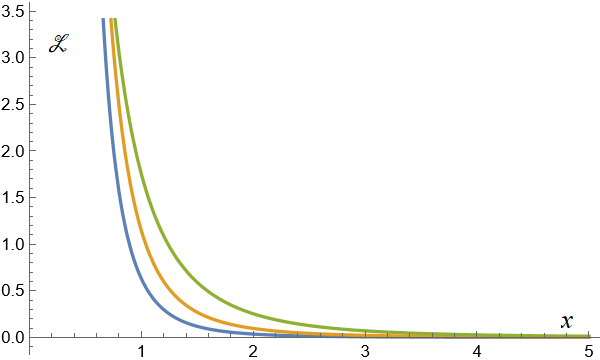}     \qq
\includegraphics[scale=0.55]{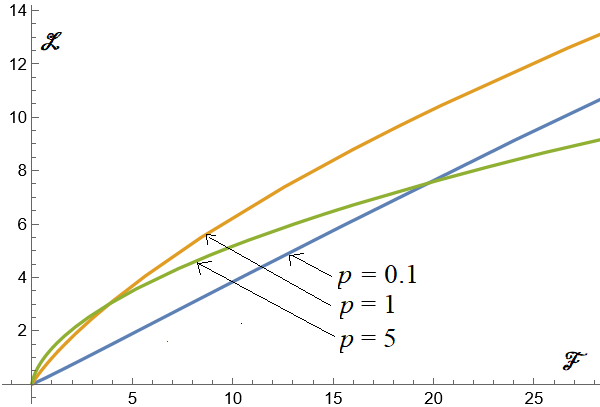}     \\
              (a)   \hspace{2in}   (b)  
\caption{\protect\small 
		 A possible source of the regularized solution for a dilatonic \bh:
		 (a) The function $\cL(\cF(x))$ for $k = 1, c =0.2$, and $p=0.1,\ 1,\ 5$ (bottiom-up).
		 (b) The function $\cL(\cF)$ for the same parameter values.
		}
\end{figure*}                                           

  The function $h(\phi(x))$ is, as before, easily calculated from \eqn{01}, where we again choose 
  $\phi = \arctan{x/c}$. Then, with the functions $A=\Areg(x)$ and $r= \rreg(x)$ given by \rf{Areg-D} 
   and \rf{rreg-D}, we obtain 
\bearr    
		h(\phi(x)) = -\frac{r''}{r} \frac{(x^2+c^2)^2}{c^2}
		= - \frac{pu}{4 c^2 (p + u)^2 (c^2 + 2 c p + p^2 + x^2)^2}
		\Big\{	-6 c^6 + 12 c^5 u 
\nnnv		\qq
		+ 12 c^4 (2 p^2 + 2 p u - x^2) +  p x^2 (4 p^3 - 3 p^2 u - 6 p x^2 - 3 u x^2) + 
       4 c^3 (6 p^3 + 3 p^2 u - 3 p x^2 + 2 u x^2) 
\nnnv       \qq
       +   c^2 (6 p^4 + 10 p^2 x^2 + 9 p u x^2 - 6 x^4) + 
       4 c (4 p^3 x^2 - 3 p x^4 - u x^4)\Big\}
\ear  
  where $u = \sqrt{x^2 +c^2}$. As was the case with Fisher's metric, it happens that 
  the sign of $h(\phi)$ changes from negative to positive at growing $x$, indicating a phantom nature 
  of the scalar field $\phi$ in a strong field region near the center and its canonical nature 
  elsewhere, see Fig.\,6(a). 
\begin{figure*} \centering   
\includegraphics[scale=0.5]{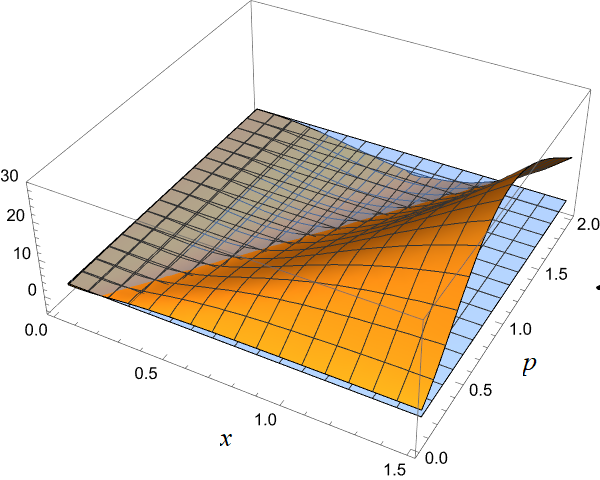}   \qq        
\includegraphics[scale=0.55]{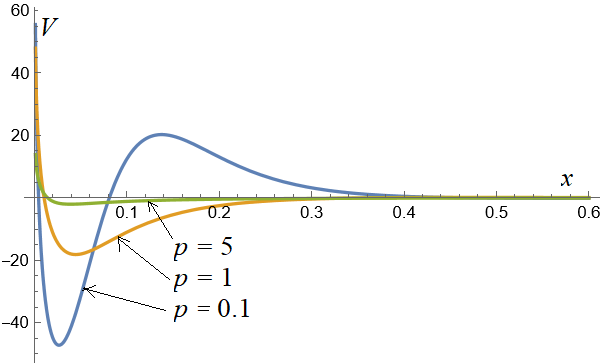}      \\
              (a)   \hspace{2in}   (b)  
\caption{\protect\small 
		A possible source of the regularized solution for a dilatonic \bh:
		 (a) The function $h(\phi(x))$ at $k=1,\ c=0.1$, and different values of $p$. 
		Its behavior is qualitatively similar to Fig.\,3(a).		
		(b)  The function $V(\phi(x))$ for $k=1,\ c=0.2,\ p = 0.1, 1, 5$. 
		}                      
\end{figure*}                                           
  The scalar field potential  $V(\phi(x)$, determined using \eqn{L(x)}, is shown in Fig.\,6(b); 
  as before, its $\phi$ dependence is easily found by substituting $x = c \tan \phi$.

\section{Concluding remarks}

\begin{enumerate}
\item  
  In the present paper, extending the study begun in \cite{kb24}, we discuss the ways of singularity removal 
  applicable to some kinds of \ssph\ space-times, by properly changing a neighborhood of a singularity 
  at $r=0$ so that it becomes a regular center. It is each time achieved by introducing a small regularization
  parameter, but quite different algorithms are required when dealing with different kinds of singularities. 
  Thus, in all cases where the original singular space-time satisfying the condition $R^t_t = R^x_x$, 
  the Bardeen-like regularization proposed in \cite{kb24} preserves this condition, and the resulting 
  space-time can be presented as a magnetic solution of Einstein-NED equations, similarly to papers 
  describing regular \bhs\ with a NED source (see \cite{kb01-NED,kb22-NED} and references therein).
\item   
  More involved are cases where the above condition is not fulfilled, and here we considered singularities 
  described by powers of the quasiglobal coordinate $x$ that does not coincide with the spherical 
  radius $r$. The simplest examples of such space-times are given by Fisher's solution for a massless minimally 
  coupled scalar field in GR and some known dilatonic \bh\ solutions, which are here subject to singularity
  removal. Quite evidently, the metrics that are better represented with other coordinates, can be 
  regularized by other (but similar) algorithms. On the other hand, even if it is hard to transform a
  particular metric to the quasiglobal coordinate $x$, quite probably such a transition can be carried
  out approximately near the singularity, and then the algorithms described here will be applicable.           
\item  
  It has been previously shown \cite{kb22} that any \ssph\ metric may be described as a solution 
  of GR with a combined source consisting of a magnetic field obeying NED and a scalar field with a
  certain self-interaction potential. It, however, turned out that such a description of the presently regularized 
  Fisher and dilatonic \bh\ solution leads to integrals that cannot be found analytically, whereas the same 
  problem for their black-bounce regularizations was solved much easier \cite{kb22}, although the global
  structure  of  space-time is more complicated in the latter case. 
\item
  All black-bounce space-times contain regular minima of the spherical radius $r$ and thus inevitably require 
  phantom matter as a source if considered as solutions of GR. Obtaining regular centers instead of 
  singularities does not necessarily require such exotic matter, though does not always avoid it.
  Thus, the examples considered in \cite{kb24} with $R^t_t = R^r_r$ did not require any exotic matter; 
  unlike that, in the presently considered examples of regularized Fisher and dilatonic \bh\ 
  metrics it turned out that such exotic matter is needed near the center, as shown in Figs.\,3(a) and 6(a).    
\end{enumerate}  
  
   One can further study various features of the new regular metrics obtained here as well as those
   resulting from other similar regularizations, in particular, their geodesic structures, gravitational lensing,
   quasinormal modes, stability, thermodynamic properties, etc. In particular, the stability properties of 
   these metrics may crucially 
   depend on the dynamics of their sources of gravity. Thus, for example, it is known that the stability 
   of the Ellis wormhole geometry \cite{ellis73, kb73} depends on whether its source is a phantom scalar 
   field, a perfect fluid or a k-essence field \cite{gonz08, kb-sha13, kb-fa21}. On the other hand, Fisher's 
   solution \rf{fish} was found to be unstable due to its behavior near its naked singularity \cite{kb-hod}, 
   and it would be of interest to know, how its different regularizations (the black-bounce one \cite{kb22} 
   and the one described here) and their associated sources affect its stability. 
   
\subsection*{Funding}
   
    We acknowledge the support from RUDN Project no. FSSF-2023-0003.
  

\end{document}